\documentclass[12pt]{elsarticle}
\usepackage[utf8]{inputenc}
\usepackage{graphicx}
\usepackage{subfigure}

\usepackage{amssymb}
\usepackage{amsthm}
\usepackage{url}
\usepackage{amsmath}
\usepackage{subfigure}
\usepackage{textcomp}
\usepackage{xcolor}

\graphicspath{{figs/}}

\journal{BioSystems}

\begin{document}

\begin{frontmatter}

\title{Fungal electronics}

\author{
Andrew Adamatzky$^{1,*}$, 
Phil Ayres$^{2}$,
Alexander E. Beasley$^{3}$,
Alessandro Chiolerio$^{1,4}$,
Mohammad M. Dehshibi$^{5}$,
Antoni Gandia$^{6}$,
Elena Albergati$^{7,8}$,
Richard Mayne$^{1}$,
Anna Nikolaidou$^{1}$,
Nic Roberts$^{1}$,
Martin Tegelaar$^{9}$,
Michail-Antisthenis Tsompanas$^{1}$,
Neil Phillips$^{1}$,
Han A. B. W\"{o}sten$^{9}$}

\address{%
$^{1}$ \quad Unconventional Computing Laboratory, UWE, Bristol, UK; \\
$^{2}$ \quad The Centre for Information Technology and Architecture, Royal Danish Academy, Copenhagen, Denmark;\\
$^{3}$ \quad Centre for Engineering Research, University of Hertfordshire, UK;\\
$^{4}$ \quad  Center for Bioinspired Soft Robotics, Istituto Italiano di Tecnologia, Via Morego 30, 10163 Genova, Italy;\\
$^{5}$ \quad Department of Computer Science, Multimedia and Telecommunications, Universitat Oberta de Catalunya, Barcelona, Spain;\\
$^{6}$ \quad Institute for Plant Molecular and Cell Biology, CSIC-UPV, Valencia, Spain;\\ 
$^{7}$ \quad Department of Design, Politecnico di Milano, Milan, Italy;\\
$^{8}$ \quad MOGU S.r.l., Inarzo, Italy;\\
$^{9}$ \quad  Microbiology, Department of Biology, University of Utrecht, Utrecht, The Netherlands;\\
}


\begin{abstract}
Fungal electronics is a family of living electronic devices made of mycelium bound composites or pure mycelium. Fungal electronic devices are capable of changing their impedance and generating spikes of electrical potential in response to external control parameters. Fungal electronics can be embedded into fungal materials and wearables or used as stand alone sensing and computing devices. 
\end{abstract}


\end{frontmatter}


\section{Introduction: Why fungal electronics?}

Flexible electronics, especially electronic skins and e-textiles~\cite{soni2020soft,ma2017self,zhao2017electronic,stoppa2014}, are amongst the most rapidly growing and promising fields of novel and emergent hardware. Flexible electronic devices are made of flexible materials where electronics capable of tactile sensing~\cite{chou2015chameleon,yang2015tactile,wang2015recent,pu2017ultrastretchable}~are embedded. Flexible electronic materials are capable of low level perception~\cite{chortos2016pursuing,park2014stretchable} and could be developed as autonomous adaptive devices~\cite{nunez2019energy}. Typical designs of flexible electronic devices include thin-film transistors and pressure sensors integrated in a plastic substrate~\cite{wang2013user}, micro‐patterned polydimethylsiloxane with carbon nanotube ultra-thin films~\cite{wang2014silk,sekitani2012stretchable}, a large-area film synthesised by sulfurisation of a tungsten film~\cite{guo2017transparent}, multilayered graphene~\cite{qiao2018multilayer}, platinum ribbons~\cite{zhao2017electronic}, polyethylene terephthalate based silver electrodes~\cite{zhao2015flexible}, digitally printed hybrid electrodes for electromyographic recording ~\cite{scalisi2015} or for piezoresistive pressure sensing ~\cite{chiolerio2014}, or channels filled with intrinsically conductive polymers~\cite{chiolerio2020tactile}.

\begin{figure}[!tbp]
    \centering
    \includegraphics[width=0.7\textwidth]{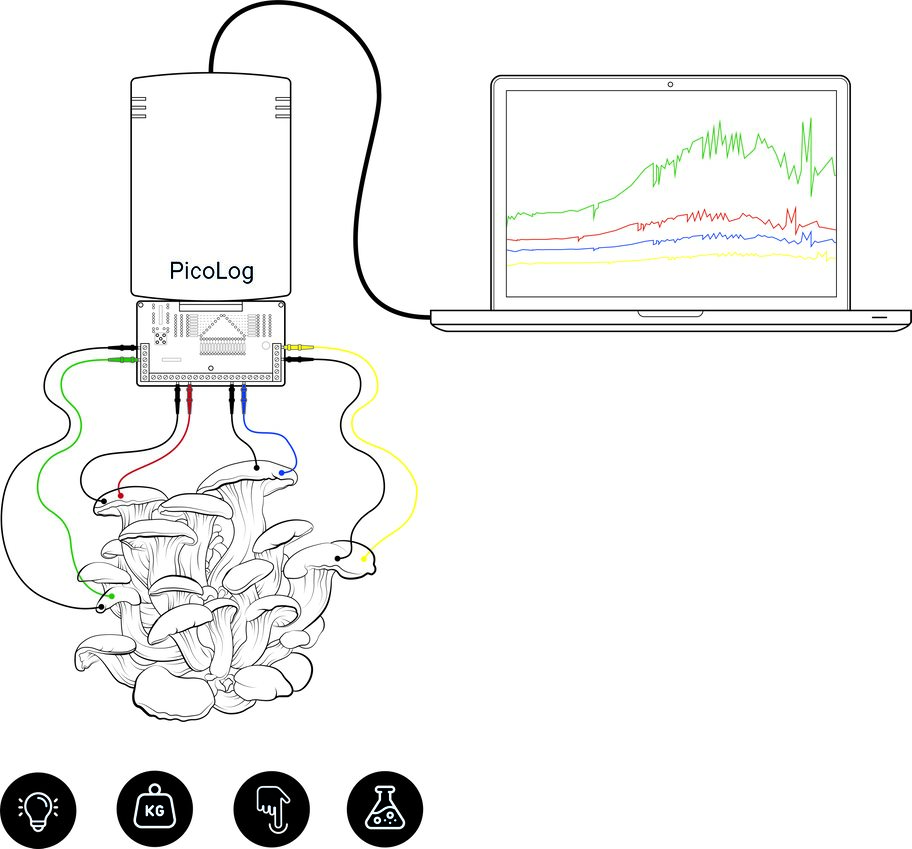}
    \caption{Caption}
    \label{fig:my_label}
\end{figure}

Whilst existing designs and implementations are highly impactful, the prototypes of flexible electronics lack a capacity to self-repair and grow. Such properties are useful, and could be necessary for organic electronics used in applications such as unconventional living architecture~\cite{adamatzky2019fungal}, soft and self-growing robots~\cite{el2018development,sadeghi2017toward,rieffel2014growing,greer2019soft} and development of intelligent materials from fungi~\cite{Meyer2020eurofung, haneef2017advanced,Jones2020composites, Wosten2019MMs} and bacteria~\cite{acetobacter2021}. Based on our previous experience with designing tactile, colour sensors from slime mould \emph{Physarum polycephalum}~\cite{adamatzky2013towards, adamatzky2013slime, whiting2014towards} and our recent results on fungal electrical activity~\cite{adamatzky2018spiking,beasley2020capacitive,beasley2020mem}, as well as following previously demonstrated thigmotropic~\cite{almeida2017thigmo}  and phototropic response~\cite{corrochano2019light} in higher fungi, we overview our recent experimental results on electronic properties of mycelium bound composites.

In this article we report that fungi exhibit properties of memristors (resistors with memory), electronic oscillators, pressure, optical and chemical sensors, and electrical analog computers. 

\section{Fungal memristors}
\label{fungalmemristors}

A memristor, also known as Resistive Switching Device (RSD), is a two or three-terminal device whose resistance depends on one or more internal state variables of the device~\cite{adamatzky2013memristor}. A memristor is defined by a state-dependent Ohm's law. Its resistance depends  on the entire past signal waveform of the applied voltage, or current, across the memristor. Using memristors, one can achieve circuit functionalities that it is not possible to establish with resistors, capacitors and inductors, therefore the memristor is of great pragmatic usefulness. Potential unique applications of memristors have been enabled by their physical implementation and are expected to occur in spintronic devices, ultra-dense information storage, neuromorphic circuits, human brain interfaces and programmable electronics~\cite{chua2019handbook,FiN2017}.

\begin{figure}[!tbp]
    \centering
\subfigure[]{\includegraphics[width=0.4\linewidth]{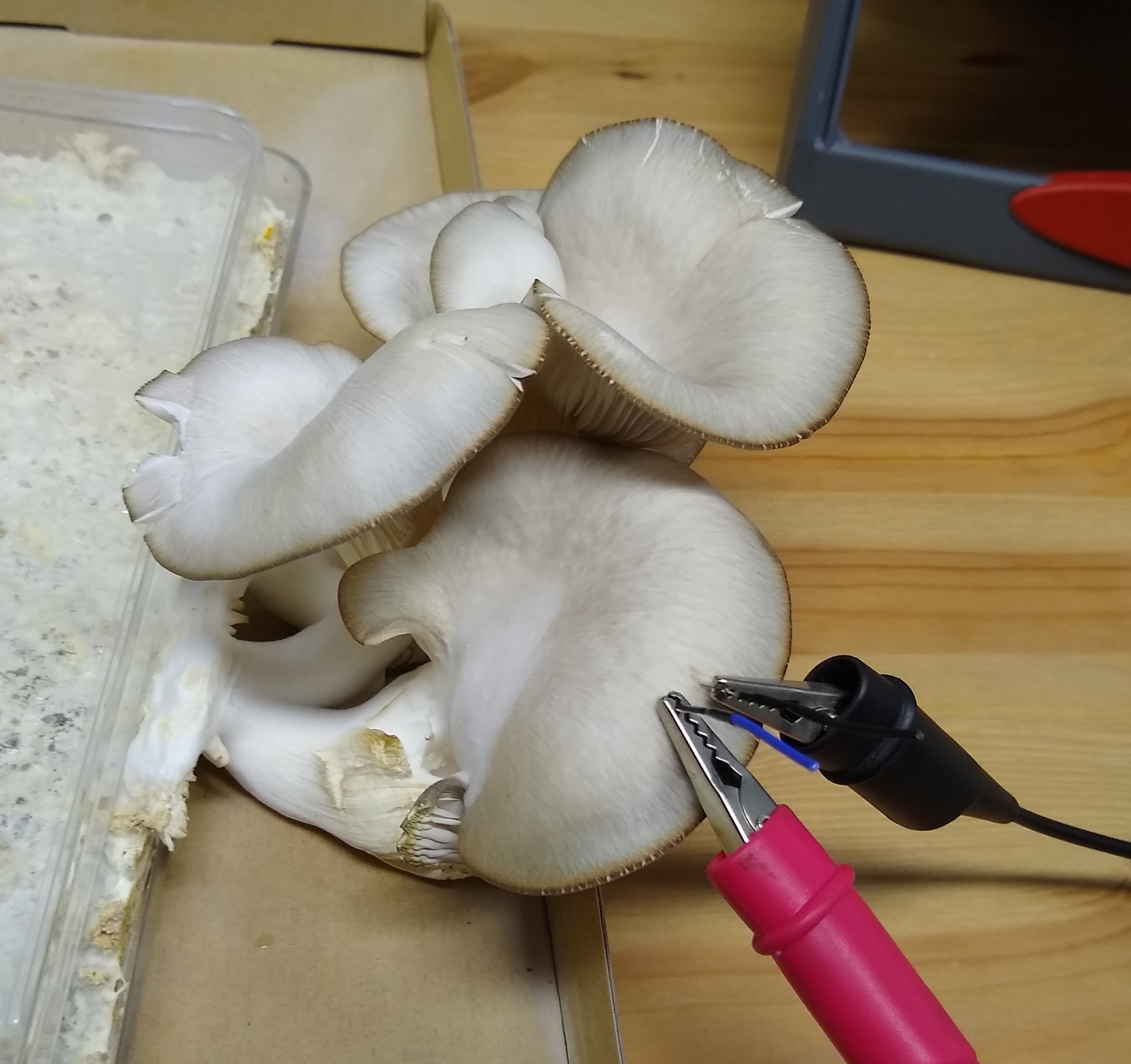}\label{fig:captocap}}
\subfigure[]{\includegraphics[width=0.4\linewidth]{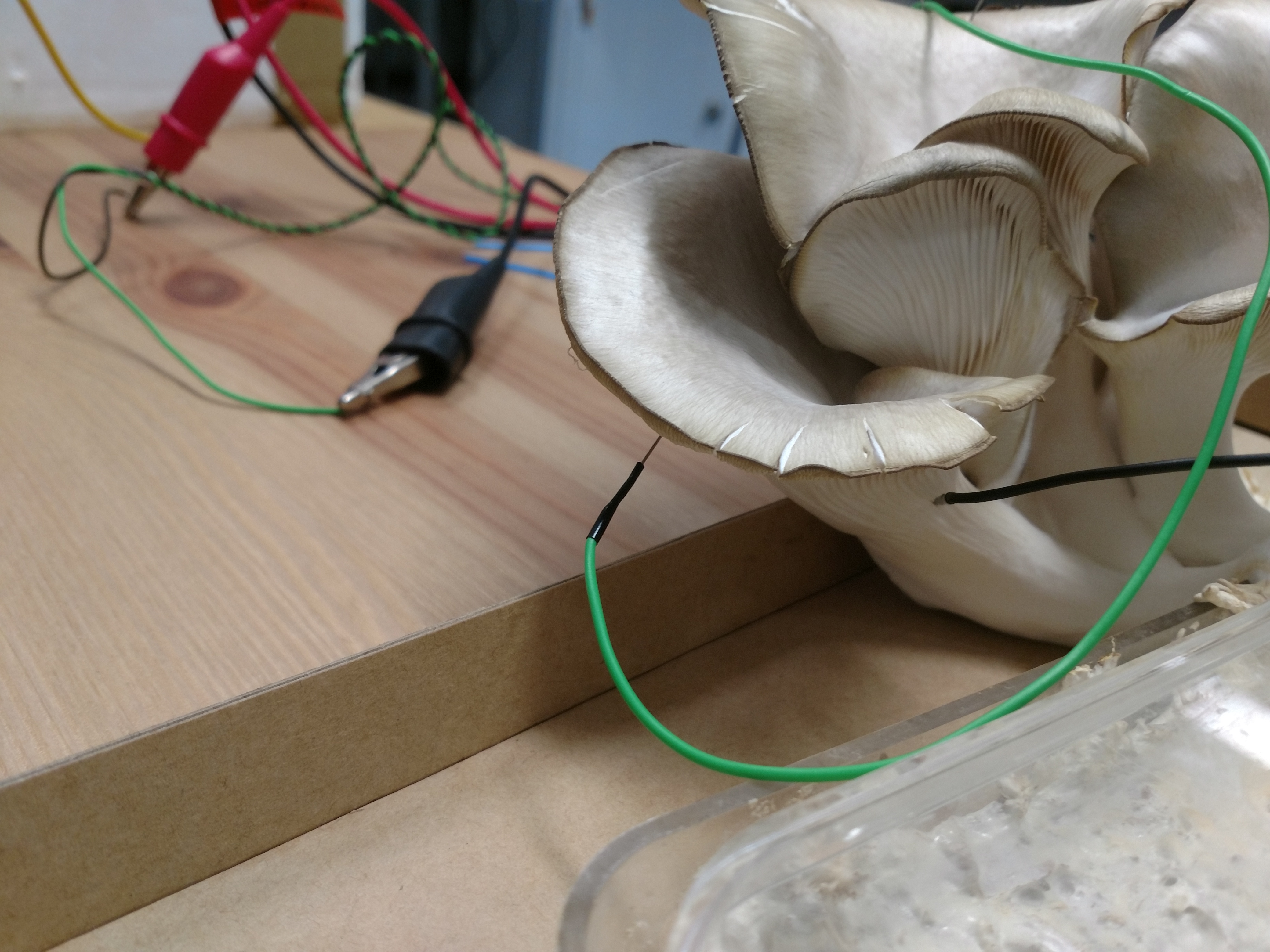}\label{fig:captostem}}
\subfigure[]{       \includegraphics[width=0.49\linewidth]{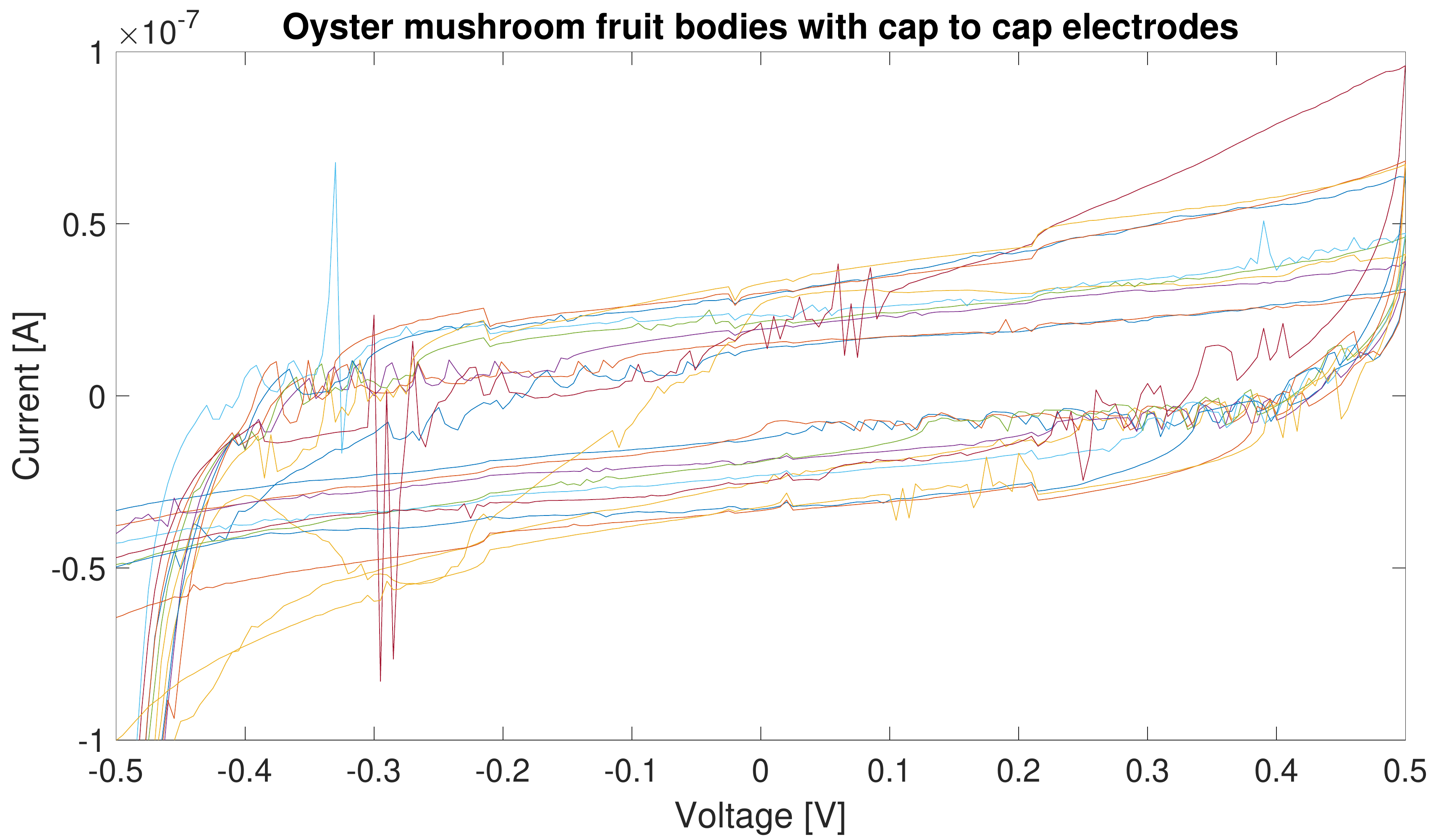}\label{cap2cap}}
\subfigure[]{
        \includegraphics[width=0.49\linewidth]{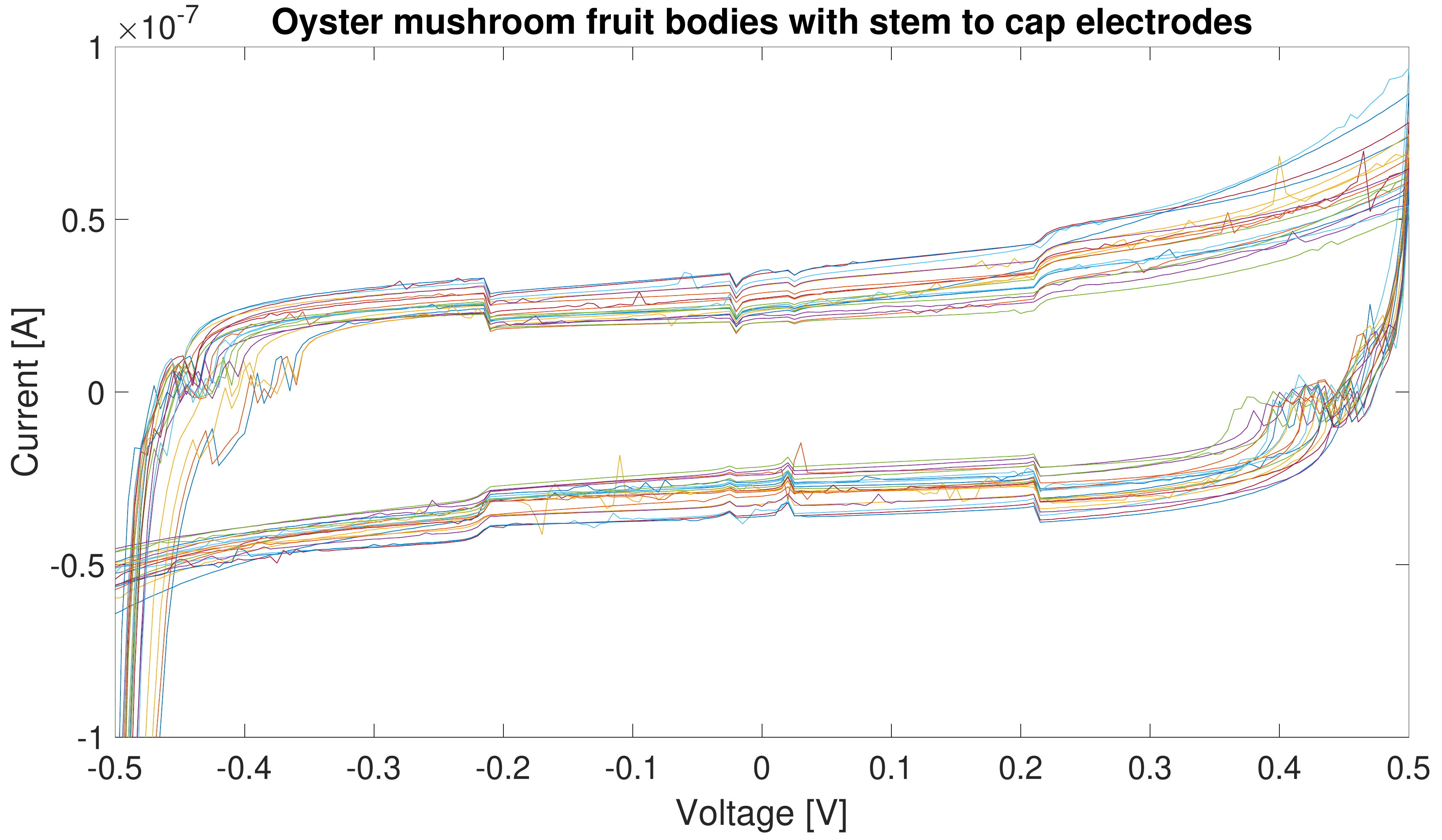}\label{stem2cap}}
    \caption{Fungal memristors. (ab)~Positions of electrodes in fruit bodies. (a)~Electrodes inserted 10~mm apart in the fruit body cap. (b)~One electrode is inserted in the cap with the other in the  stem. 
    (cd)~Raw data from cyclic voltammetry performed over -0.5~V to 0.5~V. (c)~Cap-to-cap electrode placement. (d)~Stem-to-cap electrode placement. From~\cite{beasley2020mem}. }
   \label{fig:fruitbodiesrecording}
\end{figure}

In experimental laboratory studies (see the setup in Figs.~\ref{fig:captocap} and \ref{fig:captostem}), we demonstrated that \emph{P. ostreatus} fruit bodies exhibit memristive properties when subject to a voltage sweep~\cite{beasley2020mem}. The ideal memristor model has a crossing point at 0V, where theoretically no current flows. Figures~\ref{cap2cap} and~\ref{stem2cap} show the results of cyclic voltammetry of grey oyster mushrooms with electrodes positioned in the caps and/or stems. When 0~V is applied by the source meter, a reading of a nominally small voltage and current is performed. The living membrane is capable of generating potential across the electrodes, and hence a small current is observed. To conclude, living fungi can be used as memristors (resistors with memory) in biocomputing circuits.

\section{Fungal oscillators}
\label{fungaloscillators}

\begin{figure}[!tbp]
\subfigure[]{
    \includegraphics[width=0.49\linewidth]{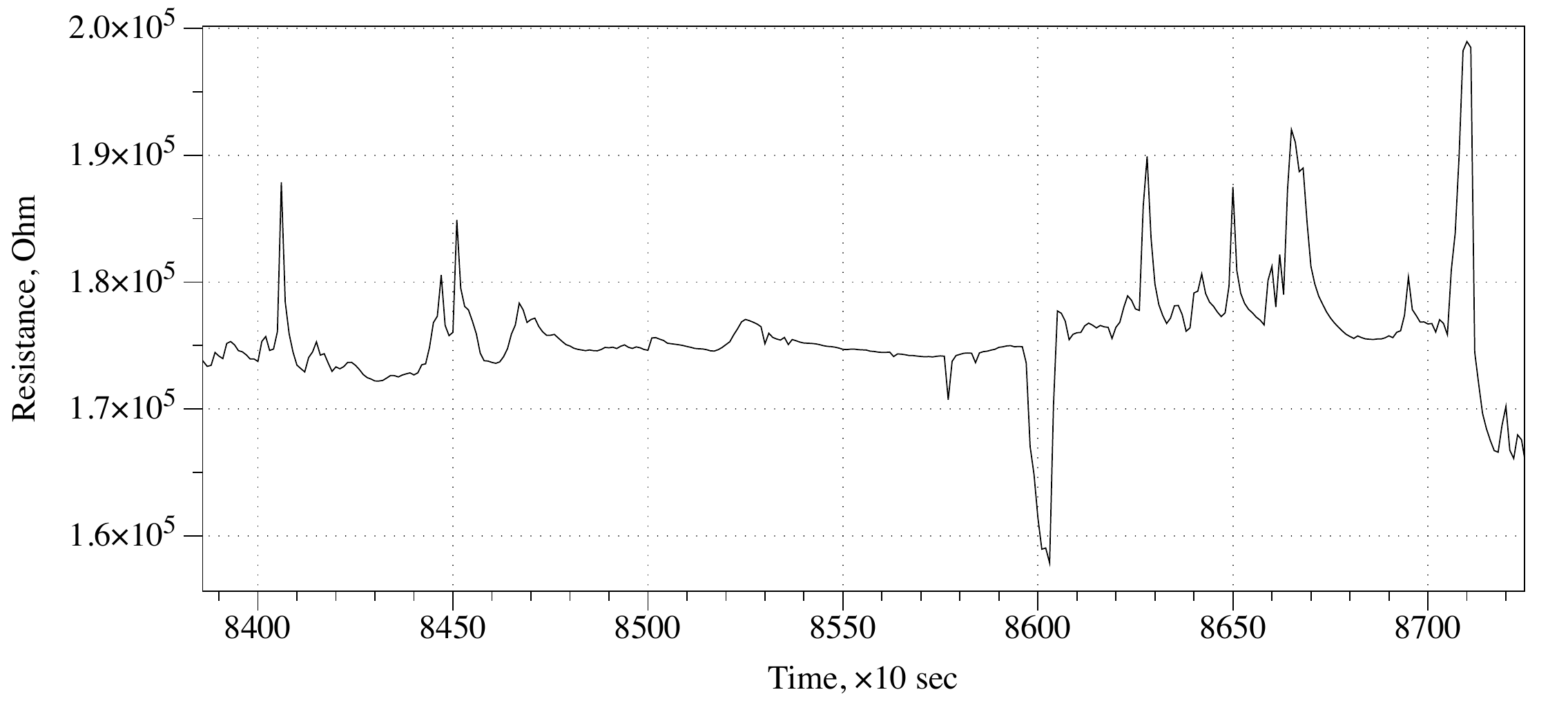}\label{fig:HighFrequency}
    }
     \subfigure[]{
    \includegraphics[width=0.49\linewidth]{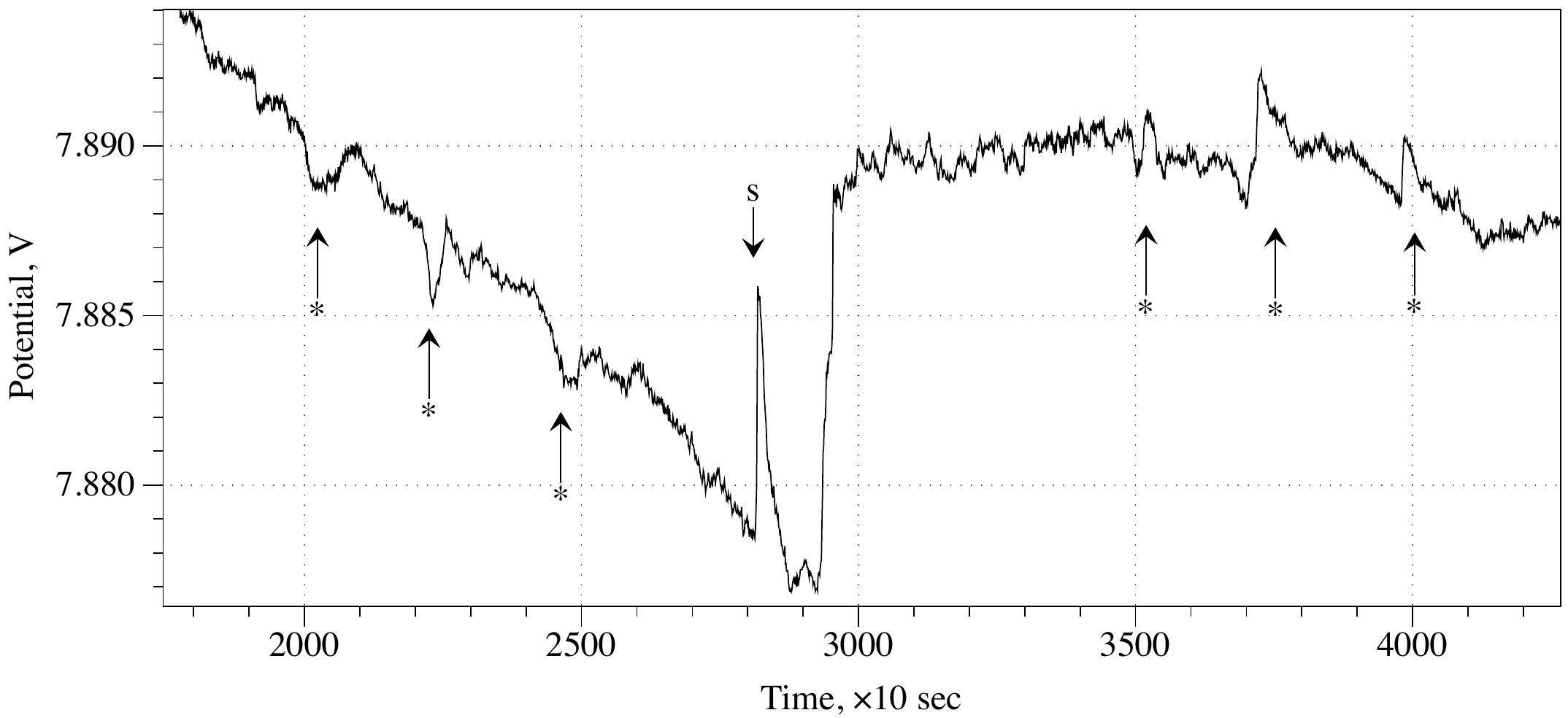}\label{fig:voltageSpikes}}
    \caption{
    (a)~Examples of high amplitude and high frequency spikes.
(b)~Oscillation of electrical potential under 10~V DC applied, where spikes analysed are marked by `*'. From~\cite{adamatzky2011electrical}.}
    \label{fig:examplsOfResistanceSpikes}
\end{figure}

An electronic oscillator is a device which converts direct current to an alternating current signal. A fungal oscillator is based on endogenous oscillations of an electrical resistance of mycelium bound composites. 
A nearly homogeneous sheet of mycelium of \emph{P. ostreatus}, grown on  the surface of a growth substrate, exhibits trains of resistance spikes (Fig.~\ref{fig:HighFrequency})~\cite{adamatzky2011electrical}. The average width of spikes is c.~23~min and the average amplitude is c.~1~k$\Omega$. The distance between neighbouring spikes in a train of spikes is c.~30~min. Typically there are 4-6 spikes in a train of spikes. Two types of electrical resistance spikes trains are found in fruit bodies: low frequency and high amplitude (28~min spike width, 1.6~k$\Omega$ amplitude, 57~min distance between spikes) and high frequency and low amplitude (10~min width, 0.6~k$\Omega$ amplitude, 44~min distance between spikes). 
To assess feasibility of the living fungal oscillator, we conducted a series of scoping experiments by applying direct voltage to the fungal substrate and measuring output voltage. An example of the electrical potential of a substrate colonised by fungi under 10~V applied is shown in Fig.~\ref{fig:voltageSpikes}. Voltage spikes are clearly observed. Spikes with amplitude above 1~mV, marked by `*', except the spike marked by `s' have been analysed. We can see two trains of three spikes each. Average width of the spikes is 10\textsuperscript{3}~sec, average amplitude 2.5~mV, while average distance between spikes is c. 2 $\cdot$ 10\textsuperscript{3}K~sec. 
To conclude, fungi can be used as a very low frequency electronic oscillators in designs of biological circuits.

\section{Fungal pressure sensor}

\begin{figure}[!tbp]
    \centering
 \subfigure[]{\includegraphics[width=0.29\linewidth]{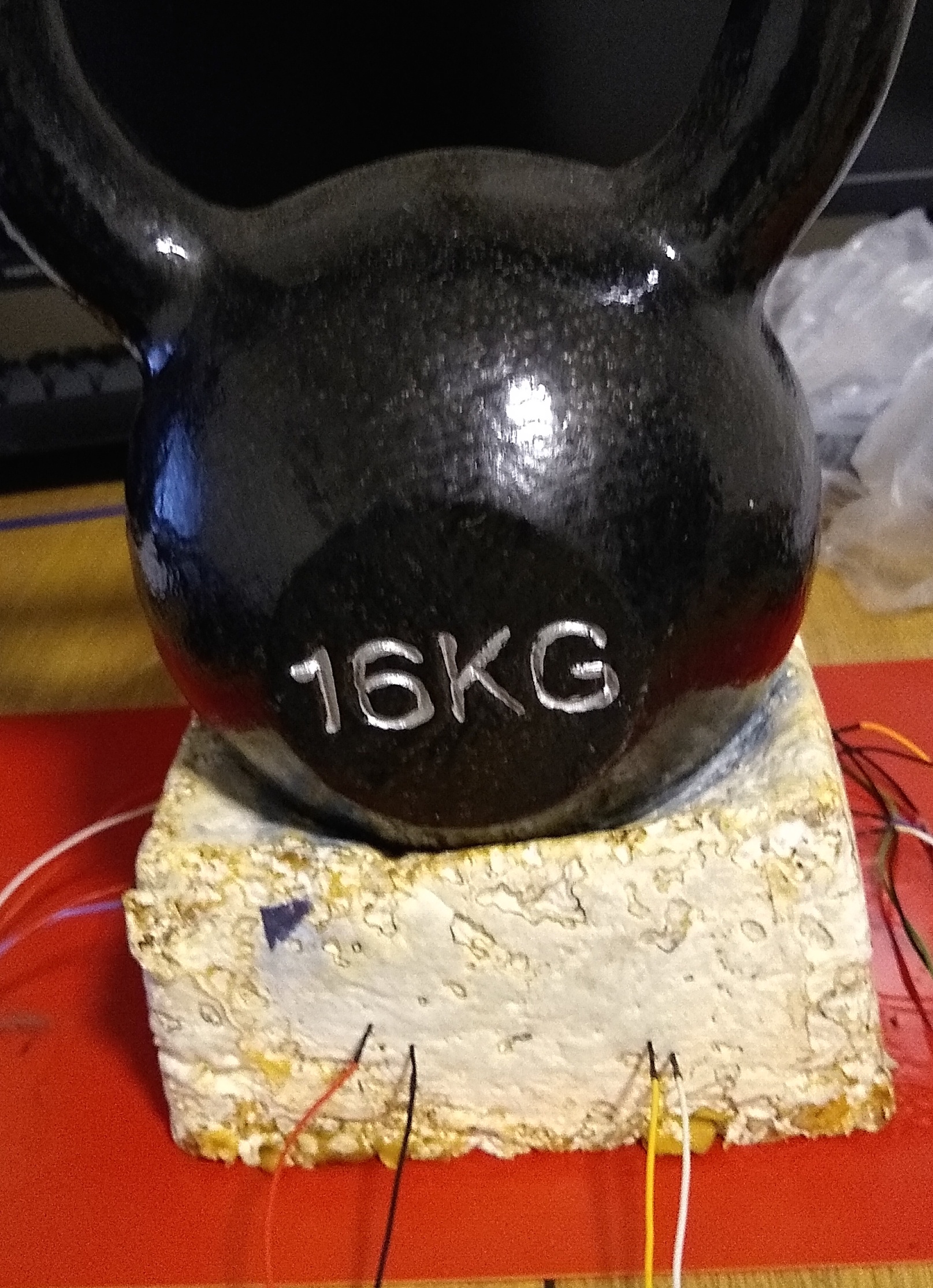}\label{16kgA}}
   \subfigure[]{\includegraphics[width=0.99\linewidth]{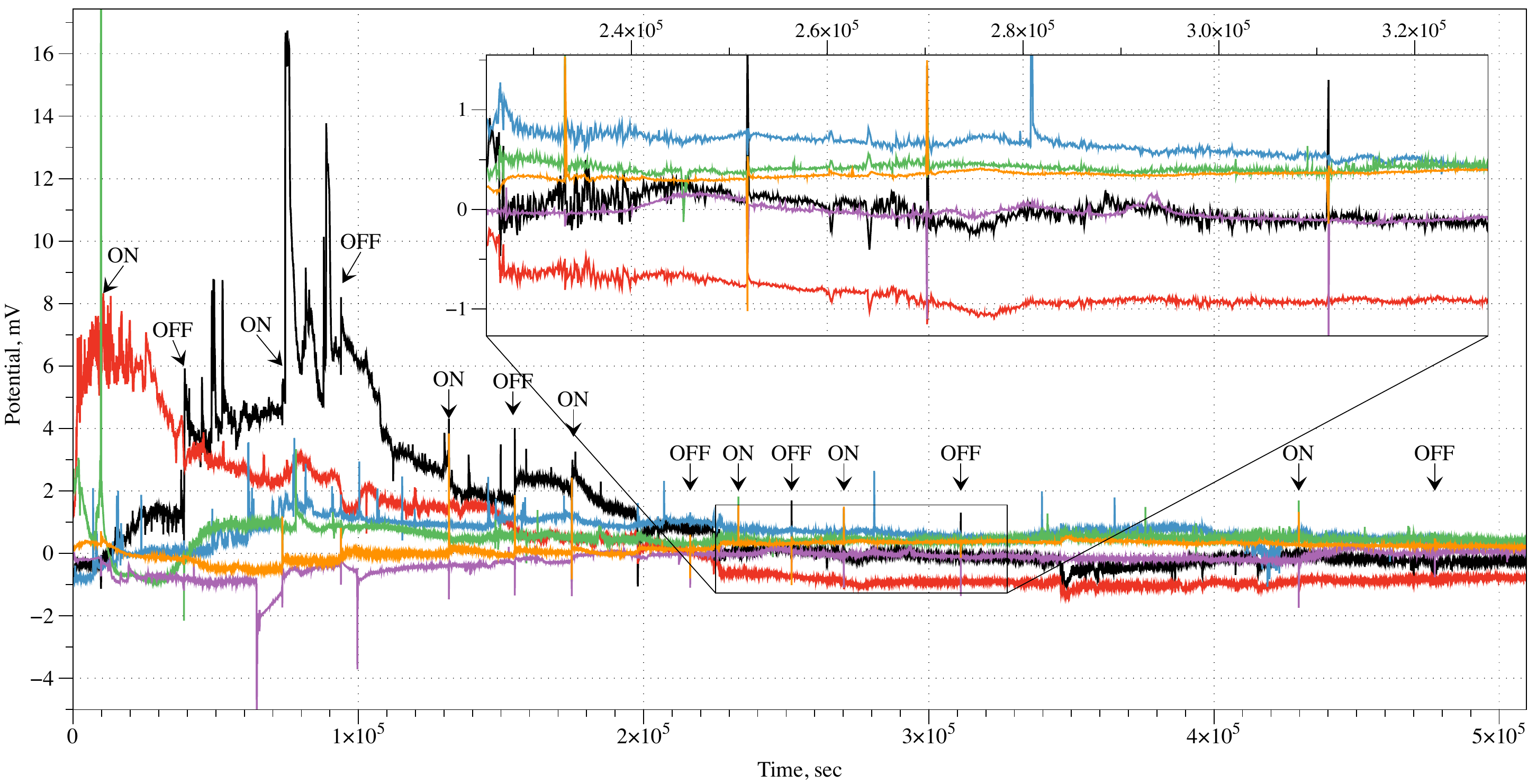}\label{4thexpAcopyA}}
    \caption{
    (a)~Experimental setup. Pairs of differential electrodes inserted in a fungal block and 16~kg kettle bell placed on top of the fungal block. Channels are from the top right clockwise (1-2), (3-5), \ldots, (15-16).
    (b)~~The activity of the block stimulated with 16~kg load.  Moments of the loads applications are labelled by `ON' and lifting the loads by `OFF'. Channels are colour coded as 
    (1-2) -- black,
    (3-4) -- red,
    (5-6) -- blue,
    (7-8) -- green,
    (9-10) -- magenta,
    (11-12)-- orange,
    (13-14) -- yellow.
    From~\cite{adamatzky2021living}.}
    \label{fig:Weigthsetup}
\end{figure}

We stimulated blocks of \emph{G. resinaceum} mycelium colonised substrate by placing a 16~kg cast iron weight on their top face (Fig.~\ref{16kgA}). Electrical activity of the fungal composite block was recorded using 8 pairs of differential electrodes, as specified in Fig.~\ref{16kgA}. 
An example of electrical activity recorded on 8 channels, during the stimulation with 16~kg weight, is shown in Fig.~\ref{4thexpAcopyA}~\cite{adamatzky2021living}. In response to application of 16~kg weight the fungal blocks produced spikes with median amplitude 1.4~mV and median duration 456~sec; average amplitude of ON spikes was 2.9~mV and average duration 880~sec. OFF spikes were characterised by median amplitude 1~mV and median duration 216~sec; average amplitude 2.1~mV and average duration 453~sec. ON spikes are 1.4 higher than and twice as longer as OFF spikes. Based on this comparison of the response spikes we can claim that fungal blocks recognise when a weight was applied or removed~\cite{adamatzky2021living}. The results complement our studies on tactile stimulation of fungal skin (mycelium sheet with no substrate) \cite{adamatzky2021fungal}: the fungal skin responds to application and removal of pressure with spikes of electrical potential. 
The fungal blocks can discern whether a weight was applied or removed because the blocks react to the application of weights with higher amplitude and longer duration spikes than the spikes responding to the removal of the weights. The fungal responses to stimulation show habituation. This is in accordance with previous studies on stimulation of plants, fungi, bacteria, and protists~\cite{applewhite1975learning,fukasawa2020ecological,ginsburg2021evolutionary,boussard2019memory,yokochi1926investigation}. To conclude, mycelium bound composites are capable of detecting pressure, therefore fungal pressure sensors can be incorporated into living loci of fungal building materials.

\section{Fungal photosensor}

\begin{figure}[!tbp]
\centering
  \subfigure[]{
    \includegraphics[width=0.45\linewidth]{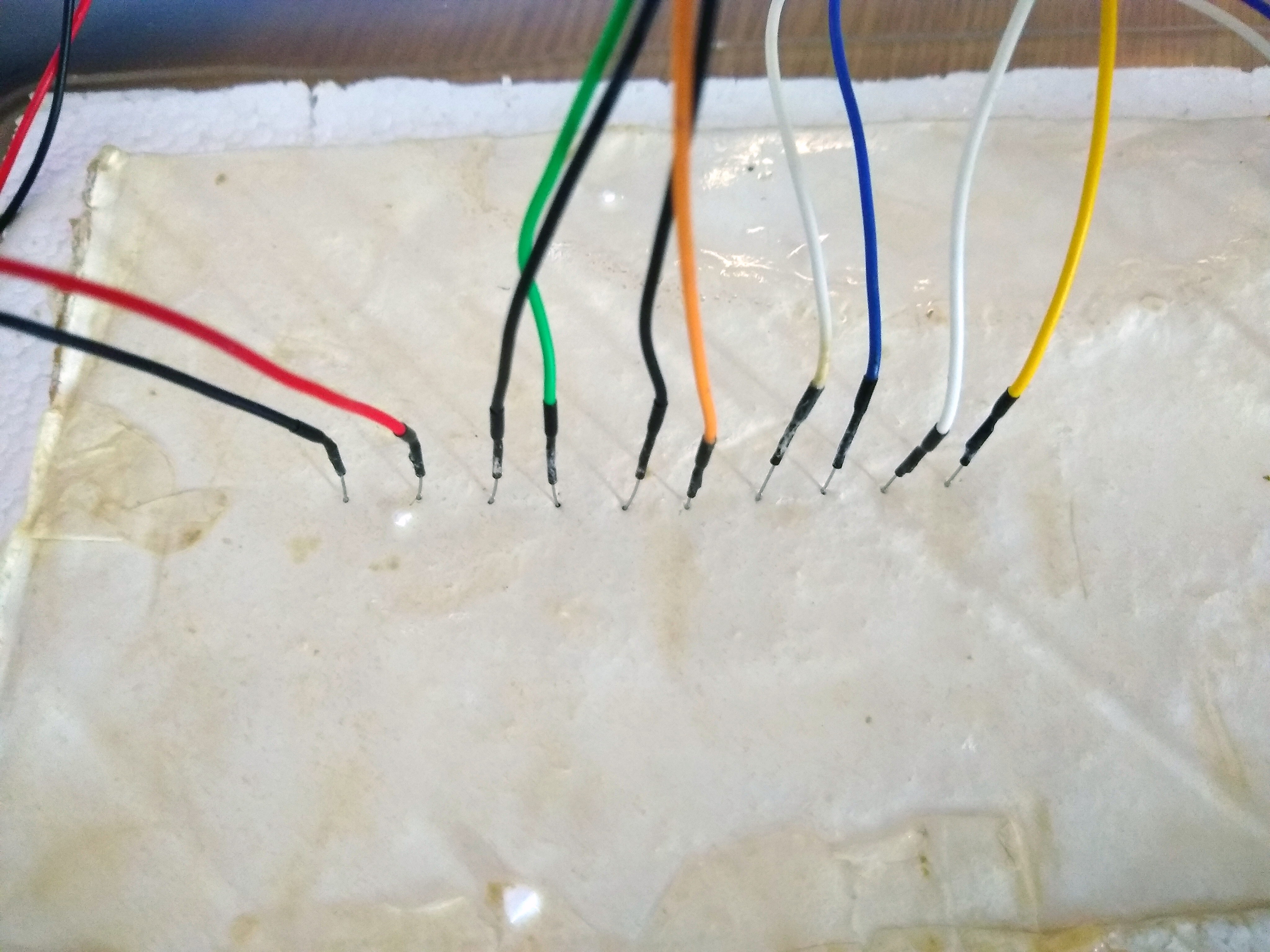}
    \label{skinphoto}
    }
    \subfigure[]{
    \includegraphics[width=0.45\linewidth]{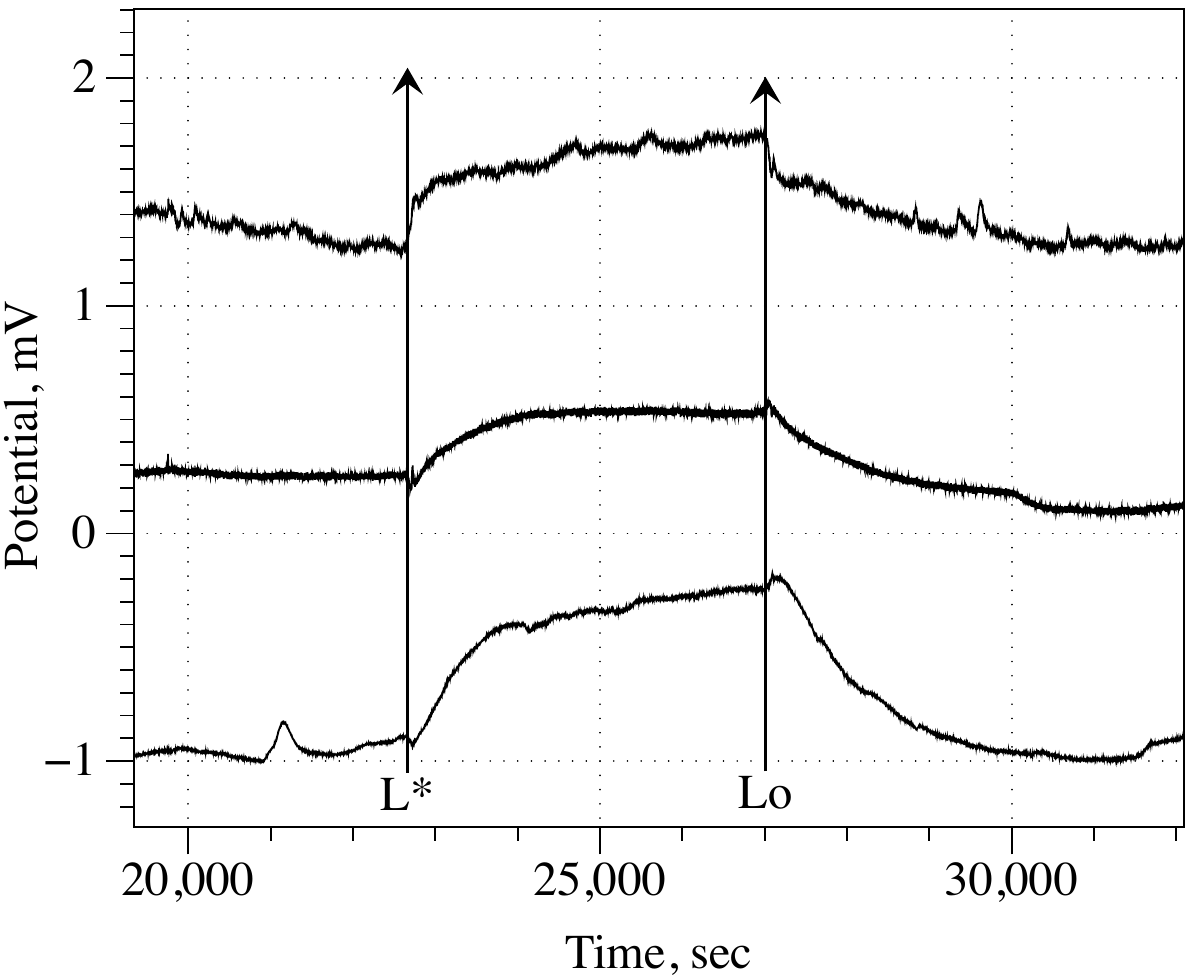}
    \label{responsetoillumination}
    }
    \subfigure[]{
    \includegraphics[width=0.45\linewidth]{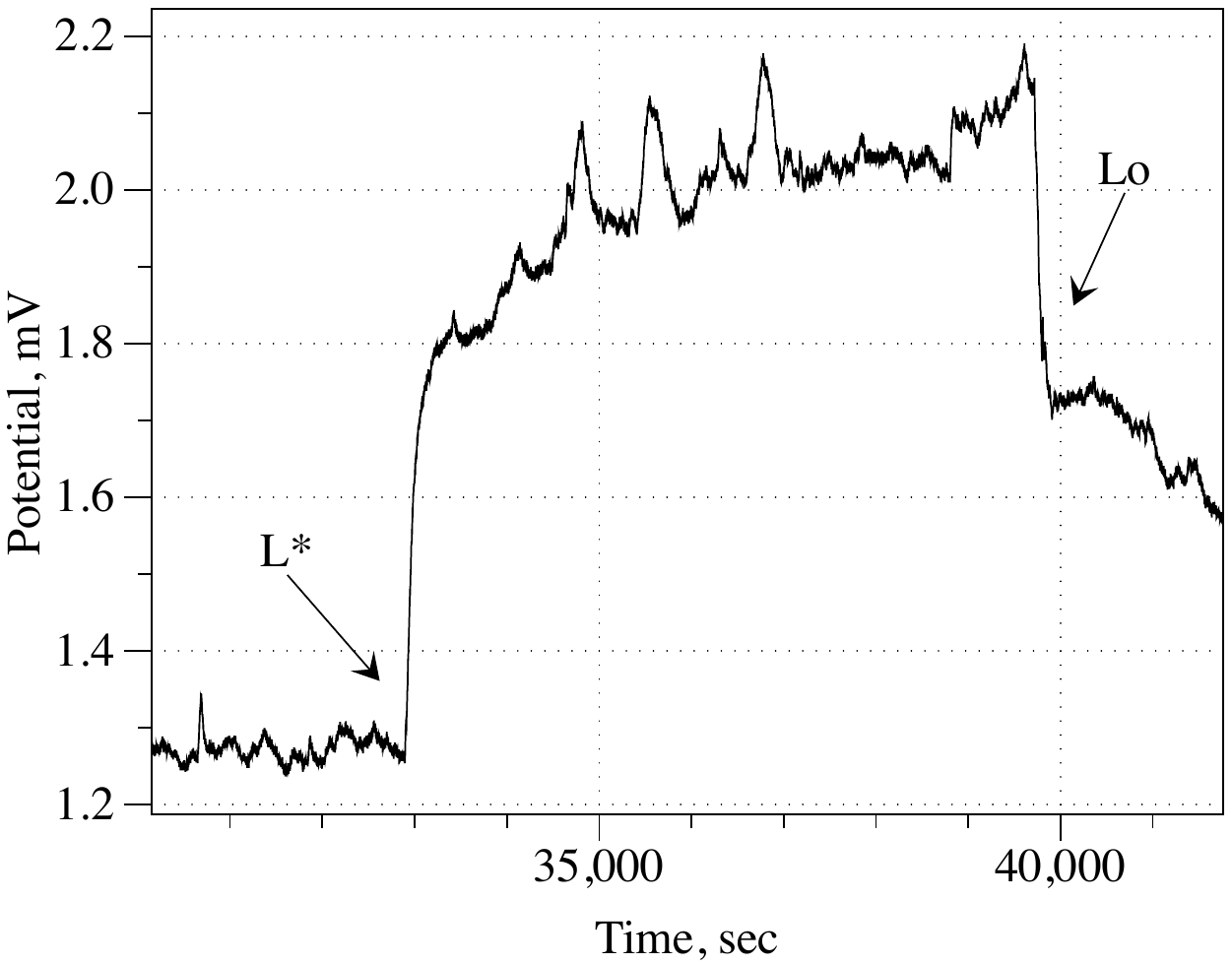}
    \label{spikesontopofresponse}
    }
    \caption{Fungal response to optical stimulation.
    (a)~A photograph of electrodes inserted into the fungal skin.
    (b)~Exemplar response of fungal skin to illumination, recorded on three pairs of differential electrodes. `L*' indicates illumination is applied, `Lo' illumination is switched off. 
    ()~A train of spikes on the raised potential as a response to illumination.
    From ~\cite{adamatzky2021fungal}.
    }
    \label{fig:examplarresponses}
\end{figure}

Fungal response to illumination was analysed using a fungal skin --- a 1.5 mm thick sheet of pure mycelium of \emph{G. resinaceum} fungi (Fig.~\ref{skinphoto})~\cite{adamatzky2021fungal}.   The response of the fungal skin to illumination is manifested in the raising of the baseline potential, as illustrated in the exemplar recordings in Fig.~\ref{responsetoillumination}. 
 The response-to-illumination spike does not subside but the electrical potential stays raised until illumination is switched off. An average amplitude of the response is 0.6~mV. The rise in potential starts immediately after the illumination is switched on. The potential saturation time is c. $3 \cdot 10^3$~sec on average; the potential relaxation time is c. $3 \cdot 10^3$~sec. Typically, we did not observe any spike trains after the illumination was switched off, however, in a couple of trials we witnessed spike trains on top of the raised potential, as shown in Fig.~\ref{spikesontopofresponse}. To conclude, living fungal materials respond to illumination by changing their electrical activity, therefore fungal materials can be incorporated in logical circuits and actuators with optical inputs.

\section{Fungal chemical sensor}

\begin{figure}[!htbp]
    \centering
    \subfigure[]{\includegraphics[width=0.3\linewidth]{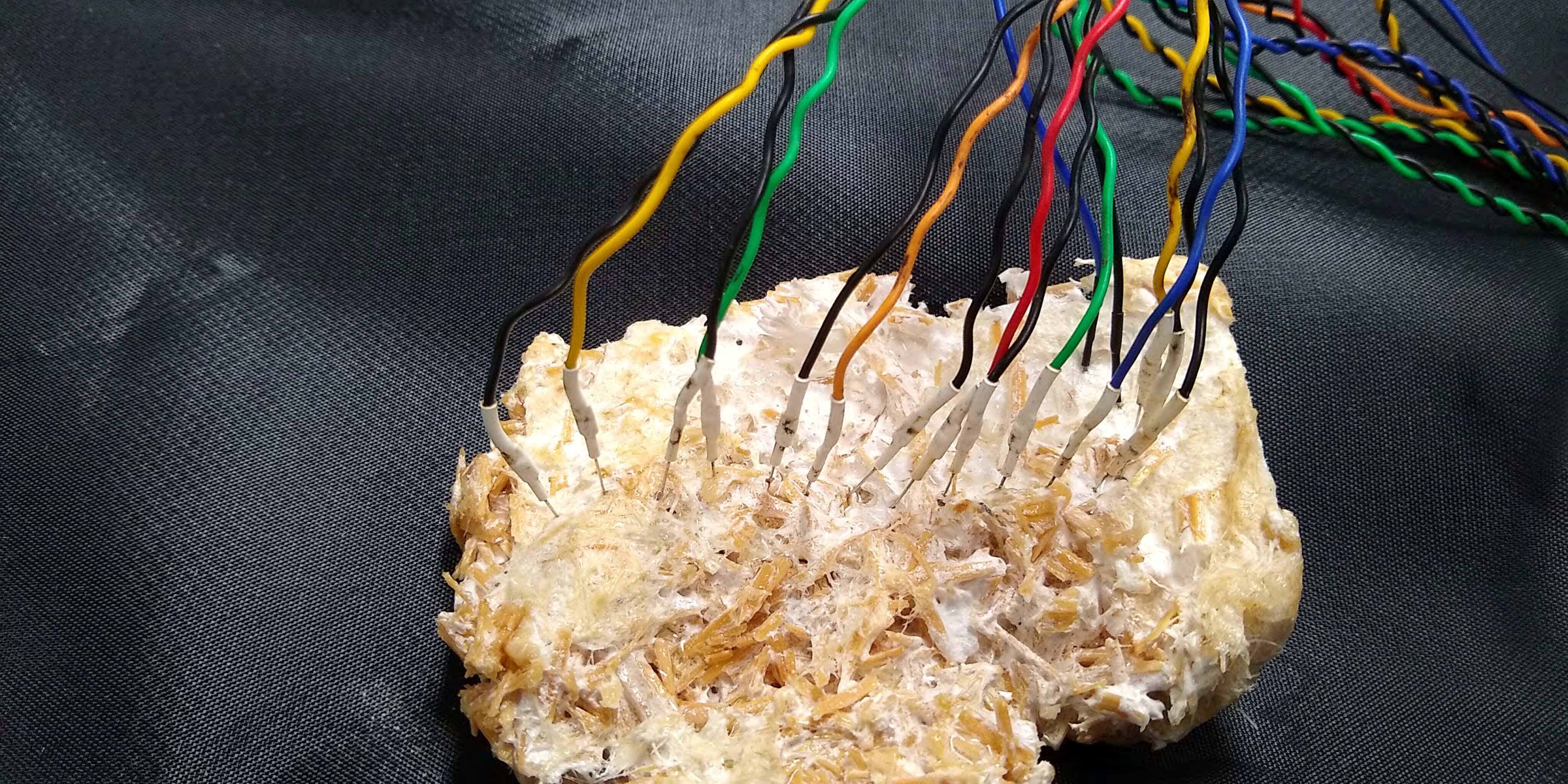}\label{fig:electrodes}}
     \subfigure[]{\includegraphics[width=0.69\linewidth]{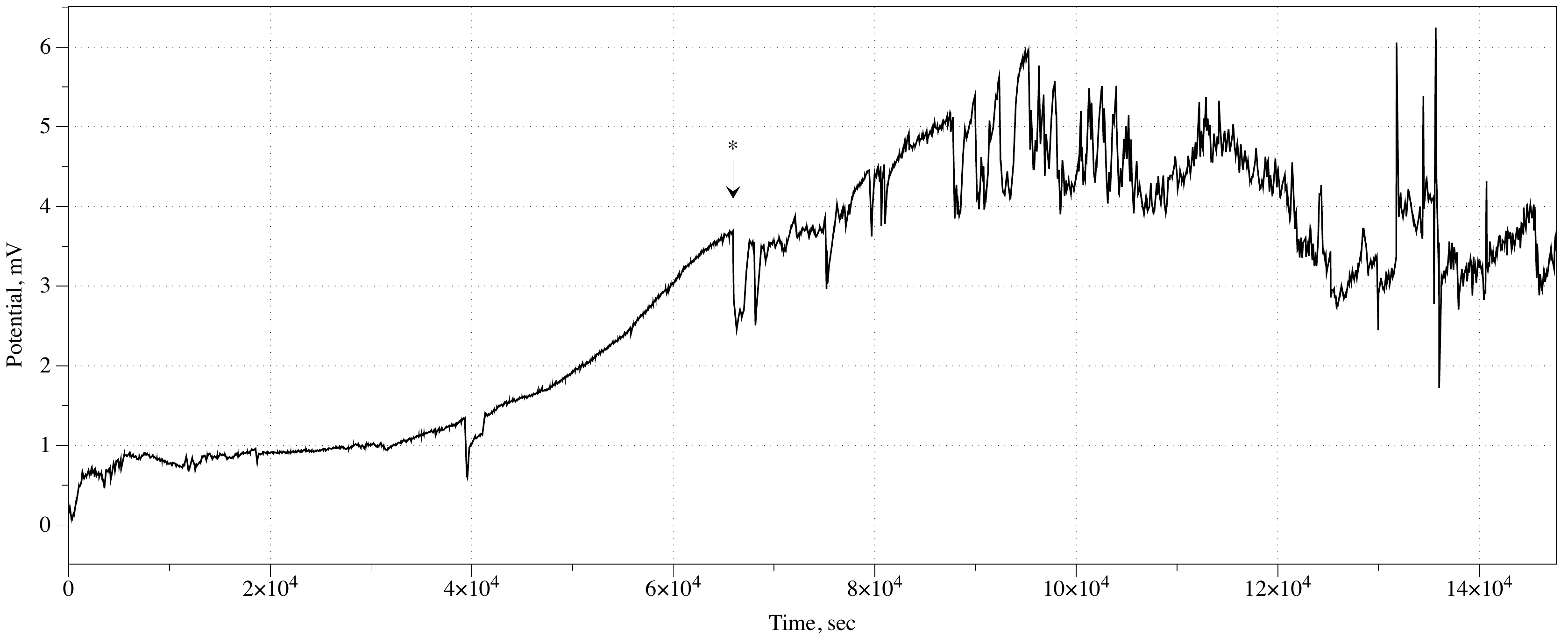}\label{fig:dextrose}}
    \caption{
    (a)~Experimental setup. Exemplar locations of electrodes.
    (b)~Response to application of dextrose. The moment of application is shown by asterisk.
    }
    \label{fig:setup}
\end{figure}

We demonstrated that hemp pads colonised by the fungus \emph{P. ostreatus} (Fig.~\ref{fig:electrodes}) show distinctive sets of responses to chemical stimulation~\cite{adamatzky2021reactive,dehshibi2021stimulating}.
We stimulated colonised hemp pads  with 96\% ethanol, malt extract powder (Sigma Aldrich, UK) dissolved in  distilled water, dextrose (Ritchie Products Ltd, UK) and hydrocortisone (Solu-Cortef trademark, 4~mL Act-O-Vial, Pfizer, Athens, Greece). An example of the response to chemical stimulation is shown in Fig.~\ref{fig:dextrose}. 
A response to stimulation with ethanol is characterised by a drop of electrical potential, up to 8~mV, followed by repolarisation phase, lasting for up to 15~sec. Fungi respond to the application of nutrients by increasing the frequency of electrical potential spiking~\cite{adamatzky2021reactive}. Exposure to hydrocortisone leads to a series of electrical disturbance events propagating along the mycelium networks with further indications of suppressed electrical activity~\cite{dehshibi2021stimulating}. 
Fungal chemical sensors show a great potential for future applications, however substantial research should be invested in their calibration.

\section{Fungal analog computing}

In numerical modelling and experimental laboratory setup we exploited principles of electrical analog computing~\cite{beasley2021electrical,roberts2021mining}.
{\sc True} and {\sc False} values are represented by above threshold and below threshold voltages. Due to the non-linearity of the conductive substrate along electrical current pathways between input and output electrodes, the input voltages are transformed and thus logical mappings are implemented. Detailed descriptions of these techniques can be found in \cite{adamatzky2020boolean,beasley2021electrical,roberts2021mining}. 

\begin{figure}[!tbp]
    \centering
    \includegraphics[width=0.5\textwidth]{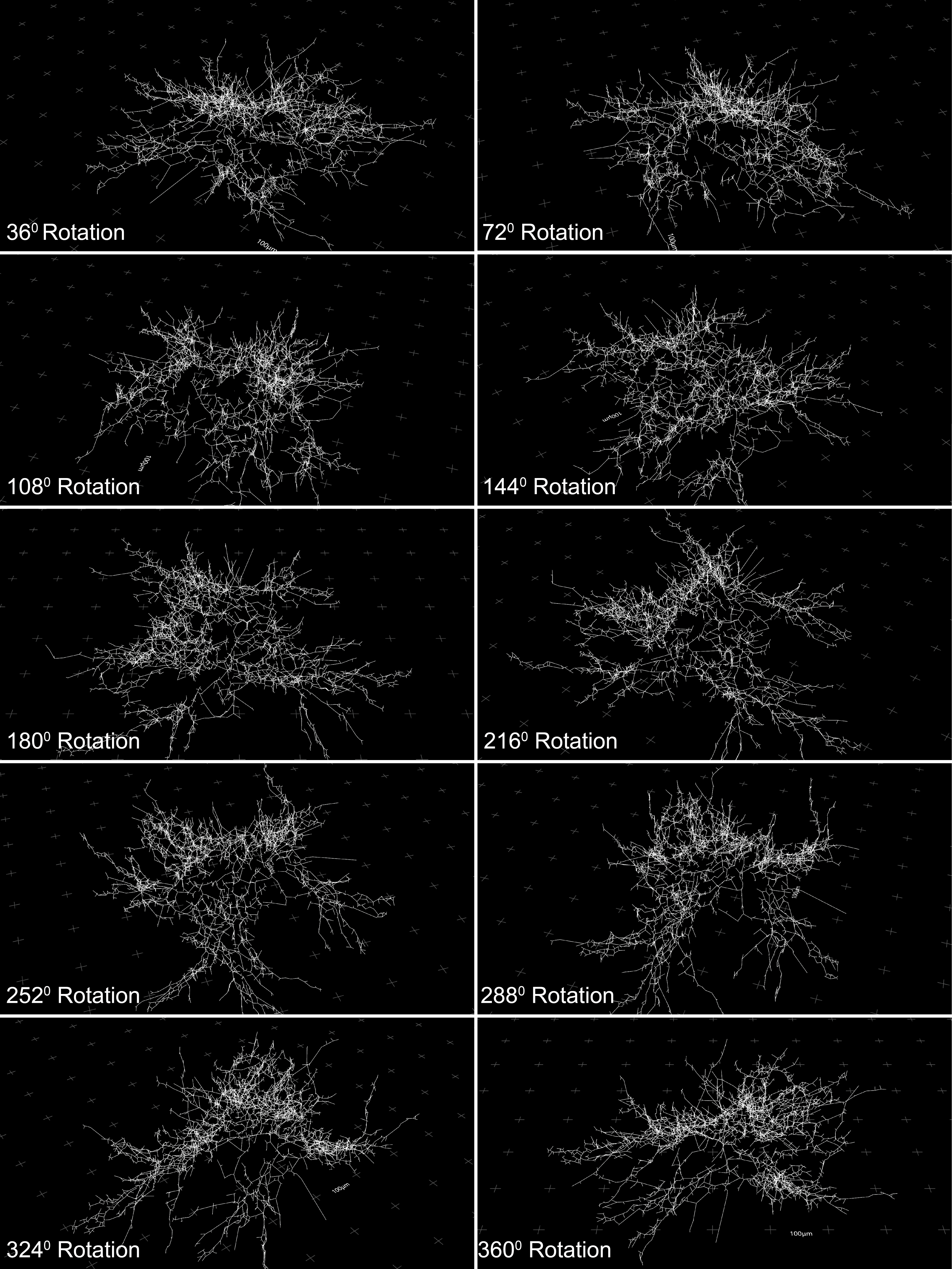}
    \caption{Perspective views of the 3D Graph. Each frame shows the graph after a 36$^{\circ}$ rotation around the z-axis with origin located approximately in the centre of the colony, on the x-y plane indicated with registration marks.}
    \label{fig:graph}
\end{figure}

The $z$-stacks of a single colony of \emph{Aspergillus niger} fungus strain  AR9\#2~\cite{vinck2011heterogenic} were converted to a 3D graph (Fig.~\ref{fig:graph}). The 3D graph was converted to a resistive and capacitive (RC) network, whose magnitudes are a function of the length of the connections. Resistances were in the order of kOhms and capacitance in the order of pF. The positive voltage and ground nodes were randomly assigned from the sample and 1000 networks were created in each arrangement for analysis. SPICE analysis consisted of transient analysis using a two voltage pulses of 60~mV on the randomly assigned positive nodes. We modelled the fungal colony in serial RC networks and parallel RC networks. The output voltages were binarised with the threshold $\theta$: $V>\theta$ symbolises logical {\sc True} otherwise {\sc False}.

\begin{figure}[!tbp]
    \centering
    \subfigure[]{\includegraphics[width=0.49\linewidth]{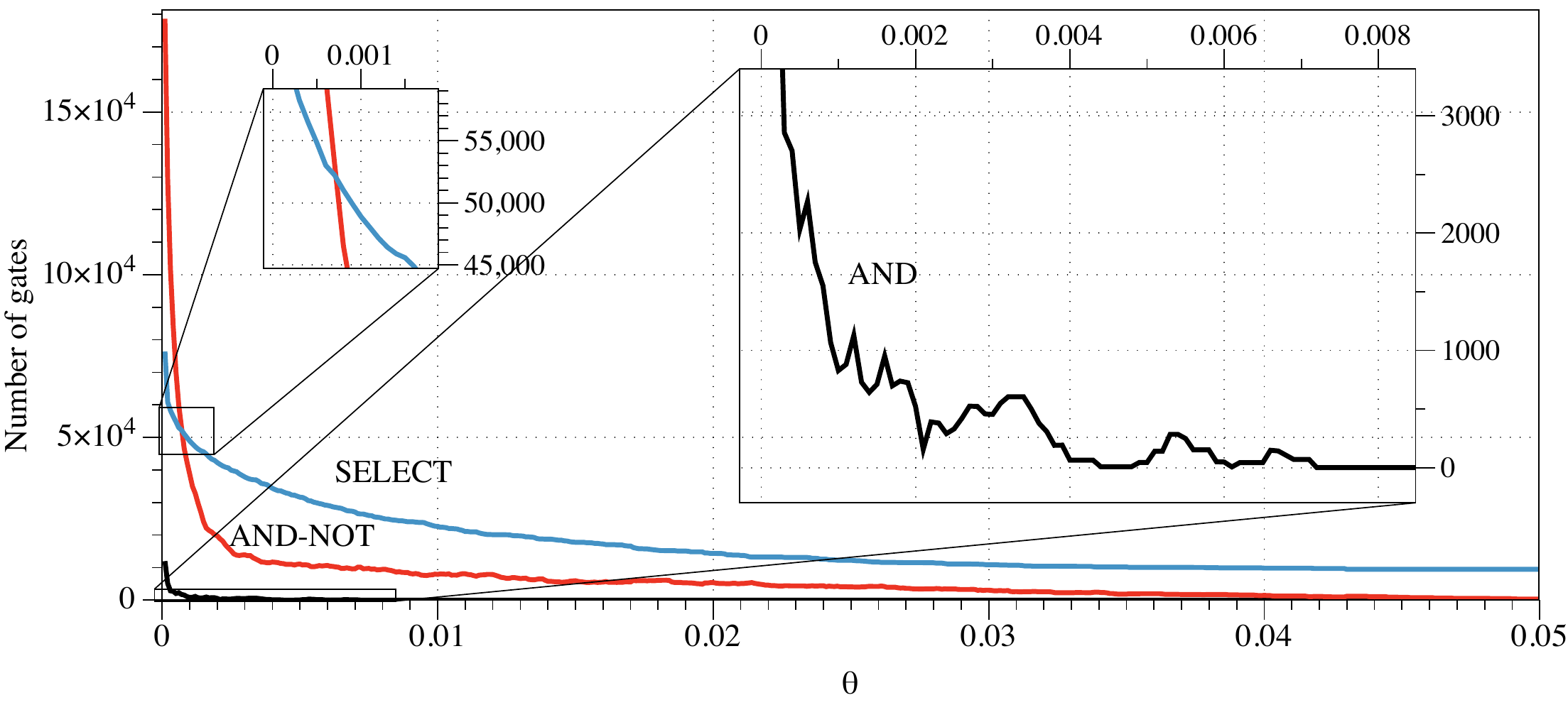}}
    \subfigure[]{\includegraphics[width=0.49\linewidth]{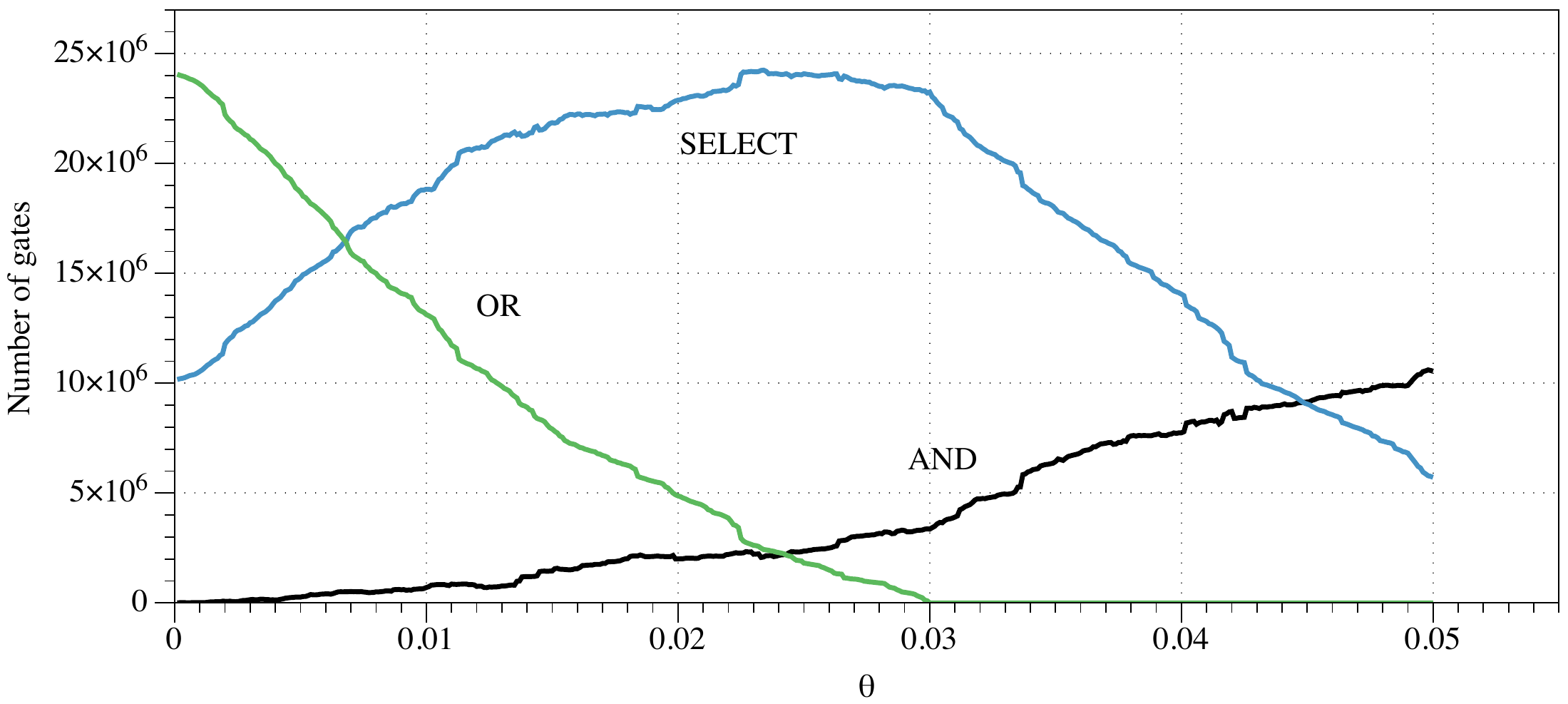}}
    \caption{Occurrences of the gates from the groups {\sc and}, black, {\sc or}, green, {\sc and-not}, red, and {\sc select}, blue, for $\theta \in [0.0001, 0.05]$, with $\theta$ increment 0.0001, in (a)~fungal colony modelled with serial RC networks, (b)~fungal colony modelled with parallel RC networks.}
    \label{fig:gates}
\end{figure}

There are 16 possible logical gates realisable for two inputs and one output. The gates implying input $0$ and evoking a response $1$, i.e. $f(0,0)=1$, are not realisable because the simulated fungal circuit is passive. The remaining 8 gates are  
{\sc and}, {\sc or}, 
{\sc and-not} ($x$ {\sc and not} $y$ and {\sc not} $x$ {\sc and} $y$), {\sc select} ({\sc select} $x$ and {\sc select} $y$) 
and {\sc xor}. In the model of serial RC networks, we found gates {\sc and}, {\sc select} and {\sc and-not}; no {\sc or} gates have been found. The number $n$ of the gates discovered decreases by a power law with increase of $\theta$. The frequency of {\sc and} gates oscillates, as shown in the zoom insert in Fig.~\ref{fig:gates}a, most likely due to its insignificant presence in the samples. The oscillations reach near zero base when $\theta$ exceeds 0.001.
In the model of parallel RC networks we only found the gates {\sc and}, {\sc select} and {\sc or}. The number of {\sc or} gates decreases quadratically and becomes nil when $\theta>0.03$. The number of {\sc and} gates increases near linearly with increase of $\theta$. The number of {\sc select} gates reaches its maximum at $\theta=0.023$, and then starts to decreases with the further increase of $\theta$.To conclude, mycelium bound composites can act as computing media and implement a wide range of Boolean circuits, thus opening a new perspective in biological analog and hybrid computing.

\section{Applications of fungal electronics}

We believe a potential practical implementation of the electronic properties of fungi would be in sensorial and computing circuits embedded into mycelium bound composites. For example, an approach of exploiting reservoir computing for sensing~\cite{athanasiou2018using}, where  the information about the environment is encoded in the state of the reservoir memristive computing medium, can be employed to prototype sensing-memritive devices from living fungi.  A very low frequency of fungal electronic oscillators does not preclude us from considering inclusion of the oscillators in fully living or hybrid analog circuits embedded into fungal architectures~\cite{adamatzky2019fungal} and future specialised circuits and processors made from living fungi  functionalised with nanoparticles, as have been illustrated in prototypes of hybrid electronic devices with slime mould~\cite{whiting2016practical,walter2016hybrid,ntinas2017oscillation,adamatzky2015twenty,berzina2015hybrid}. Electrical resistance of living substrates is used to identify their morphological and physiological state~\cite{crile1922electrical,schwan1956specific,mcadams1995tissue,heroux1994monitoring,dean2008electrical}. Examples include  determination of states of organs~\cite{gersing1998impedance}, detection of decaying wood in living trees~\cite{skutt1972detection,al2006electrical}, estimation of root vigour~\cite{taper1961estimation}, the study of freeze-thaw injuries of plants~\cite{zhang1992electrical}, as well as classification of breast tissue~\cite{da2000classification}. 

\section{Discussion}

The sensing fungal composites described here, exhibit a range of advantages compared to other living sensing materials, e.g. slime mould sensors~\cite{adamatzky2013towards,adamatzky2013slime,whiting2014towards} 
electronic sensors with living cell components~\cite{kovacs2003electronic},
 chemical sensors using living taste, olfactory, and neural cells and tissues~\cite{wu2015bioanalytical} and tactile sensor from living cell culture~\cite{minzan2013toward}. The advantages are low production costs, simple maintenance and durability. A further important advantage is scalability: a living fungal material patch can be as small as few millimeters or it can be grown to several metres in size.

\section{Acknowledgements}

This research has received funding from the European Union's Horizon 2020 research and innovation programme FET OPEN ``Challenging current thinking'' under grant agreement No 858132 / project \textit{Fungal Architectures}, \url{www.fungar.eu}.




%

\end{document}